# Piezoelectric strain induced variation of the magnetic anisotropy in a high Curie temperature (Ga,Mn)As sample

A. Casiraghi[1], A. W. Rushforth[1], J. Zemen[1,2], J. A. Haigh[1], M. Wang[1], K. W. Edmonds[1], R. P. Campion[1] and B. L. Gallagher[1]

[1] School of Physics and Astronomy, University of Nottingham, University Park, Nottingham NG7 2RD, United Kingdom

[2] Institute of Physics ASCR, v.v.i., Cukrovarnická 10, 162 53 Praha 6, Czech Republic

*We show that effective electrical control of the magnetic properties in the ferromagnetic semiconductor (Ga,Mn)As is possible using the strain induced by a piezoelectric actuator even in the limit of high doping levels and high Curie temperatures, where direct electric gating is not possible. We demonstrate very large and reversible rotations of the magnetic easy axis. We compare the results obtained from magneto-transport and SQUID magnetometry measurements, extracting the dependence of the piezo-induced uniaxial magnetic anisotropy constant upon strain in both cases and detailing the limitations encountered in the latter approach.*

The ability to manipulate the magnetization orientation of a ferromagnetic material with an electric field is an important requirement for the implementation of spintronics devices in information and communication technologies [1]. In the past decade, gate voltage control of ferromagnetism has been achieved in a number of experiments involving ferromagnetic semiconductors like (In,Mn)As and (Ga,Mn)As [2-4]. Since they rely on the possibility of varying the magnetic properties via hole depletion, these experiments are only effective for ultra-thin samples with very low hole concentrations, a requirement that is typically fulfilled in as-grown samples with low Mn doping and consequently extremely low

Curie temperatures ($T_C$). The difficulty to obtain sizeable hole depletions in highly-doped ferromagnetic semiconductors has driven research in the investigation of alternative routes to the manipulation of magnetism in these systems. In particular, the use of piezoelectric actuators to mediate, via strain, the electric control of the magnetic properties in (Ga,Mn)As has been shown to be a promising approach [5-8]. This is consistent with the fact that magnetostrictive effects are known to be large in materials characterized by strong spin-orbit coupling like (Ga,Mn)As. This approach has so far been explored for temperatures up to 50 K in samples with relatively low $T_C$ [5-8]. By using a (Ga,Mn)As sample with high hole and Mn content and a $T_C$ of 180 K we are able to both investigate the effects of the piezo-induced strain on the magnetic properties of highly-doped (Ga,Mn)As and also to work at much higher temperatures than previous experiments, where larger strains can be generated by the actuator [9]. We use magneto-transport measurements to show that reversible rotations of the magnetic easy axis by roughly 80° can be achieved in this sample, thus demonstrating the clear advantage of strain mediated control over direct electric field control of magnetism in samples with high hole concentrations. We also compare the results obtained from magneto-transport measurements with those derived from magnetometry measurements [10] and we extract the dependence of the piezo-induced uniaxial magnetic anisotropy constant upon strain in both cases.

The material used was a 25 nm thick film of $(Ga_{0.88},Mn_{0.12})As$ grown by low temperature molecular beam epitaxy on a GaAs(001) substrate and buffer layers, and annealed in air for 48 hours at 180 °C [11]. Superconducting quantum interference device (SQUID) magnetometry measurements show that the magnetic easy axis lies in the plane of the film, and is determined by a competition between a cubic anisotropy favoring the [100] and [010] orientations, and a uniaxial anisotropy favoring the [1-10] orientation. At the working temperature $T$ of 150 K the uniaxial magnetic anisotropy dominates and the easy axis is along the [1-10] direction, as revealed by the magnetic hysteresis loops in Fig. 1(a). Using the Stoner-Wohlfarth model for a single magnetic domain [12] to fit to the hard axis [110] loop we can extract the in-plane "intrinsic" uniaxial $K_U$ and cubic $K_C$ anisotropy constants at 150 K.

A piece of the material was patterned by optical lithography into a Hall bar oriented along the [1-10] direction for magneto-transport measurements while another piece was left unpatterned for SQUID measurements. Following the procedure described in Ref. [9], these samples were then bonded to piezoelectric actuators with a two-component epoxy, after thinning the substrate down to 190 ± 10 μm by chemical etching. In both cases, the [1-10] direction was chosen to be along the poling direction of the actuator, although a small misalignment $\delta$ (see Fig. 1(b)) was unavoidably introduced and found to be 2° ± 0.5° for the Hall bar and 1.5° ± 0.5° for the SQUID sample. The sign of $\delta$ determines the sense of the rotation of the uniaxial easy axis induced by the strain: for the combination of cubic and uniaxial anisotropies of our sample and positive $\delta$ we expect the easy axis to rotate anticlockwise for tensile strain and clockwise for compressive strain. The uniaxial strain induced by the actuator (tensile/compressive for positive/negative voltages, respectively) was measured via four terminal resistance measurements of a strain gauge aligned along the [1-10] direction and glued on top of the (Ga,Mn)As layer of the transport sample, next to the Hall bar. According to the manufacturer's specifications, the actuators used in the magnetometry and transport experiments should generate the same stress to within 10 %.

The orientation of the easy axis in the Hall bar sample was inferred from transverse anisotropic magnetoresistance (AMR) measurements, as in Refs. [5-8]. The transverse AMR can be expressed as $R_{xy} = \Delta R \sin(2\vartheta)$ [13], where $\vartheta$ is the angle between the magnetization direction and the current direction [1-10] (see Fig. 1 (b)). As a result of the working temperature being close to $T_C$, we found a significant dependence of the saturation magnetization $M_S$, and consequently of the AMR amplitude $\Delta R$, on the magnetic field $B$. The procedure used to extract the functional form of $\Delta R(B)$ is described in the Supplementary Information. By measuring $R_{xy}$ as a function of $B$, with $B$ applied along different in-plane directions, we identified the direction of the easy axis (within ± 1°) as that for which $R_{xy}/\Delta R$ remains constant upon variation of $B$, while it changes significantly when $B$ is applied along directions away from the easy axis. In this way the easy axis was found to have already rotated by ~ -5.5° from the "intrinsic" [1-10] direction, when no voltage $V_P$ was applied to the actuator. This is due to the creation of a small in-plane tensile strain (which could not be accurately measured) on cooling down to 150 K, arising from the different thermal contraction coefficients of the piezoelectric actuator and the GaAs [5]. Upon application of a positive $V_P$ the easy axis rotated further towards the [110] direction and reached an angle of -81° for $V_P$ = +85 V, as illustrated in Fig. 2(a), while it rotated to -3° for $V_P$ = -70 V, as illustrated in Fig. 2(b). Overall, we demonstrated that a reversible rotation of the easy axis by ~ 78° can be achieved with this device.

In the SQUID magnetometry experiment we measured hysteresis loops at various $V_P$, with $B$ applied along the [1-10] direction. An initial measurement of the actuator, prior to gluing the (Ga,Mn)As film, revealed that it is strongly paramagnetic ($M/B$ ~ 7 × 10$^{-6}$ J/T$^2$) and also has a ferromagnetic component with the remnant magnetization $M_{REM} = M(B=0)$ comparable to that of the (Ga,Mn)As film. Therefore, to determine the properties of the (Ga,Mn)As sample, the magnetic hysteresis loop of the actuator was subtracted from that of the composite device. The result is shown in Fig. 2(c). The inset shows the orientation of the easy axis determined from fitting to the measured hysteresis loops. The easy axis rotates from ~ -25° for $V_P$ = 0 V to ~ -68° for $V_P$ = +85 V, a smaller amount than that observed in the magneto-transport device, probably due to the fact that the strain induced in the relatively large SQUID sample (2.5 mm X 3 mm) is inhomogeneous and on average less than the strain induced in the small Hall bar section (50 μm X 250 μm).

We modeled the dependence of the piezo-induced uniaxial anisotropy constant $K_{Up}$ on the strain $\Delta\varepsilon$ generated by the actuator by minimizing a phenomenological expression for the magnetic free energy of the system. According to the Stoner-Wohlfarth model [12], the free energy density is given by $E = -K_C/4 \sin^2(2\varphi) + K_U \sin^2\varphi - K_{Up} \sin^2(\varphi - \delta) - M_S B \cos(\varphi - \vartheta)$, where the first, second and third terms are respectively the in-plane magnetocrystalline cubic, magnetocrystalline uniaxial and piezo-induced uniaxial magnetic anisotropy contributions, the fourth term is the Zeeman contribution and the angles are defined in Fig. 1(b). The anisotropy energies $K_U$ = 144.6 J/m$^3$ ± 4 % and $K_C$ = 45 J/m$^3$ ± 25 % were obtained from the hysteresis loops of the sample before mounting on the actuator shown in Fig. 1(a). For the magneto-transport device we extracted $K_{Up}(\Delta\varepsilon)$ from the easy axis orientation using the above formula with $B$ = 0. The result is shown by the closed squares in Fig. 3, with the error bars due to the combined uncertainties in $\delta$, $K_U$, $K_C$ and the easy axis orientation. $\Delta\varepsilon$ is arbitrarily set to 0 for $V_P$ = 0 V, which results in $\Delta\varepsilon$ ~ 6.1 × 10$^{-4}$ for $V_P$ = + 85 V. Due to the large uncertainties in deriving $K_{Up}$ from the easy axis orientation for negative voltages, these data were not included in Fig.3. The best straight line through the points has a slope of ~ 28

J/m$^3$ per 10$^{-4}$ strain. For the magnetometry device we extracted $K_{Up}(\Delta\varepsilon)$ by fitting the Stoner-Wohlfarth model to the regions of each hysteresis loop away from switching phenomena, as displayed in Fig. 2(c). The result is shown by the open circles in Fig. 3, with the error bars due to the combined uncertainties in $\delta$, $K_U$ and $K_C$. The best straight line through the points has a slope of ~ 13 J/m$^3$ per 10$^{-4}$ strain. The discrepancy in the results obtained from the two methods could be due to several factors. For the magnetometry experiment, the presence of inhomogeneous strain and multidomain processes result in a reduction of the inverse magnetostrictive effect, while the magnetic impurities in the actuator lead to experimental uncertainty. It will be important for experimentalists to consider the influence of such factors when using the magnetometry technique to study inverse magnetostriction [10]. The magneto-transport experiment requires care to properly account for the dependence of $\Delta R$ on $B$ for $T$ close to $T_C$. Notwithstanding this, plus the fact that the different samples and actuators used in each experiment may possess variations in their physical properties, it appears that the magneto-transport experiment provides the more accurate measurement of the inverse magnetostriction.

In our earlier study of magnetization reorientations induced by a piezoelectric actuator [5], we compared the measured results to calculations based on the **k·p** description of the GaAs host valence band and the kinetic-exchange model of its coupling to the local Mn moments [14-15]. The predictions for the as-grown sample nominally doped to 6% Mn were in semi-quantitative agreement with the experimental observations at T = 50 K. The same modelling predicts a perpendicular-to-plane magnetic easy axis for the high hole densities [16] occurring in our sample (around one hole per Mn atom) which has never been observed experimentally. A general trend in the calculated anisotropies is the independence of the cubic anisotropy component on lattice strains and the linear dependence of the in-plane uniaxial anisotropy component on the piezo-strain. This is in agreement with the experimental behaviour we observe. As both $K_U$ and $K_C$ decrease with increasing temperature, and the applied strain enhances $K_U$ but does not affect $K_C$, the strain-enhanced $K_U$ totally dominates the energy profile allowing for the large observed magnetization rotations.

To conclude, we have shown that large and reversible piezo-induced rotations of the magnetic easy axis can be achieved in a highly-doped (Ga,Mn)As sample with a high $T_C$. This result demonstrates the clear advantage of strain mediated control over direct electric field control of magnetism in samples with high carrier concentrations. We have compared the results obtained from magneto-transport and SQUID magnetometry measurements, extracting the dependence of the piezo-induced uniaxial magnetic anisotropy constant upon strain in both cases and detailing the limitations encountered in the latter approach.

The project was funded by EU Grant nos. SemiSpinNet-215368 and NAMASTE-214499 and UK EPSRC grant number EP/H003487/1. We are thankful to Tomas Jungwirth for his useful contributions.

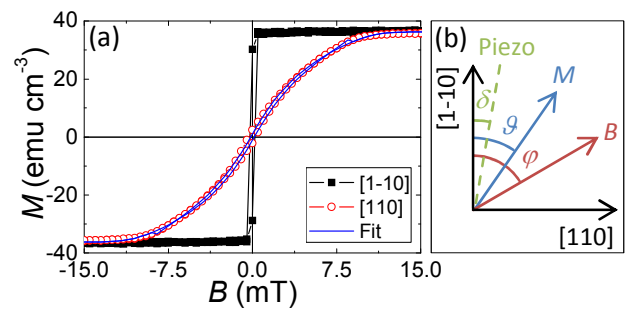

**Figure 1** (a) Magnetic hysteresis loops measured by SQUID magnetometry at $T$ = 150 K for annealed $(Ga_{0.88},Mn_{0.12})As$ with the external magnetic field $B$ applied along the [1-10] (closed squares) and [110] (open circles) directions. The solid line is the result of the fitting to the [110] loop using the Stoner-Wohlfarth model. (b) Definition of the angles mentioned in the text.

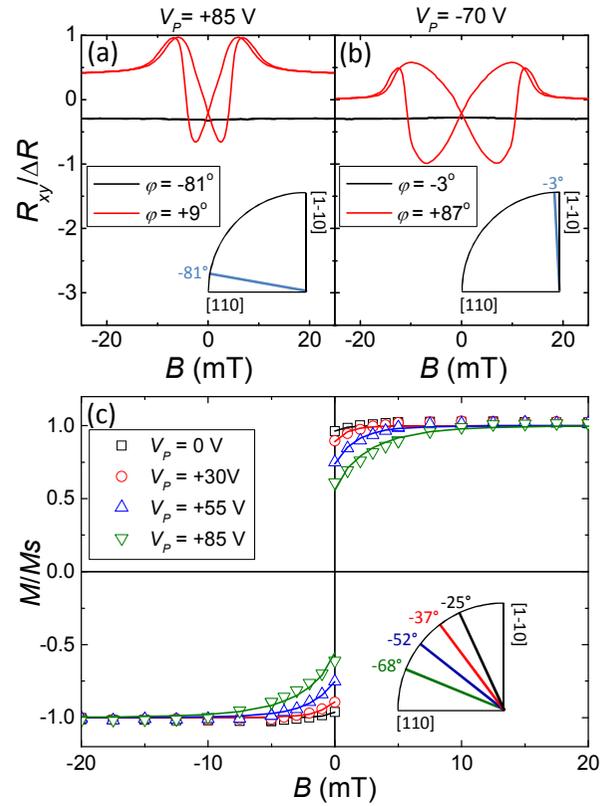

**Figure 2** Normalized transverse resistance $R_{xy}/\Delta R$ measured at $T$ = 150 K as a function of the external magnetic field $B$ applied at different in-plane angles $\varphi$ for (a) $V_P$ = +85 V and (b) $V_P$ = -70 V. (c) Magnetic hysteresis loops measured by SQUID magnetometry at $T$ = 150 K with $B$ applied along the [1-10] direction and corresponding fittings. The insets in (a), (b) and (c) show the easy axis direction for each applied piezo-voltage $V_P$.

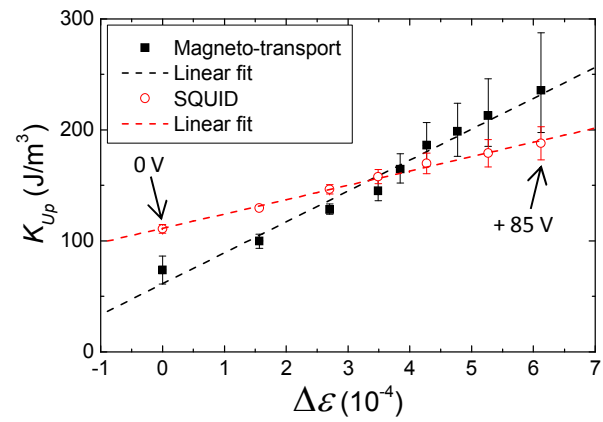

**Figure 3** Dependence of the piezo-induced uniaxial magnetic anisotropy constant $K_{Up}$ on strain $\Delta\varepsilon$ at $T$ = 150 K as derived from magneto-transport (closed squares) and SQUID magnetometry (open circles) experiments. The corresponding linear fits are shown by the dashed lines.

# Supplementary Information

**Derivation of the dependence of the amplitude of the AMR on the magnetic field**

In a ferromagnet for temperature close to $T_C$ the polarization induced by a magnetic field $B$ on the magnetic moments is non negligible and results in a dependence of the saturation magnetization $M_S$ at that temperature on $B$ [1]. Since the amplitude $\Delta R$ of the AMR is influenced by the value of $M_S$, the dependence of $M_S$ on $B$ determines a dependence of $\Delta R$ on $B$. If $\Delta R$ is proportional to $M$, as suggested in Ref. [2], then the functional expression of $M(B)$ can be used for $\Delta R(B)$. According to mean-field theory $B = a(T-T_C)M + bM^3$ [3], with $a$ and $b$ positive constants. This means that $M$, and consequently $\Delta R$, can be expressed as: $M(B) \sim \Delta R(B) \sim a'B + b'B^{1/3}$.

Within an approximation of $\sim 10\%$ the transverse AMR is given by $R_{xy} = \Delta R \sin(2\vartheta)$ [4], where $\vartheta$ is the angle between the direction of $M$ and the current direction [1-10] in the Hall bar. Fig. S1(a) shows $R_{xy}$ as a function of $B$ when $B$ is applied in the plane of the Hall bar at the angles $\varphi = -45°$ and $\varphi = +45°$ to the [1-10] direction. For saturating $B$ (for which $\varphi = \vartheta$), these angles are those at which $R_{xy}$ is respectively maximum and minimum ($\Delta R < 0$), as displayed by the inset in Fig. S1(a). Therefore, for saturating $B$, $\Delta R(B)$ can be simply derived by taking the difference between the two curves in Fig. S1(a) and dividing it by 2. $\Delta R(B)$ thus obtained is shown for $B > 45$ mT by the square symbols in Fig. S1(b), along with the mean field curve used to model it: $\Delta R(B) = \Delta R_0 + a'B + b'B^{1/3}$, where $\Delta R_0 = -0.575$, $a' = -0.029$ and $b' = -0.312$.

The expression of $\Delta R(B)$ thus derived for $T = 150$ K is the one used for the analysis of the magneto-transport data described in the paper.

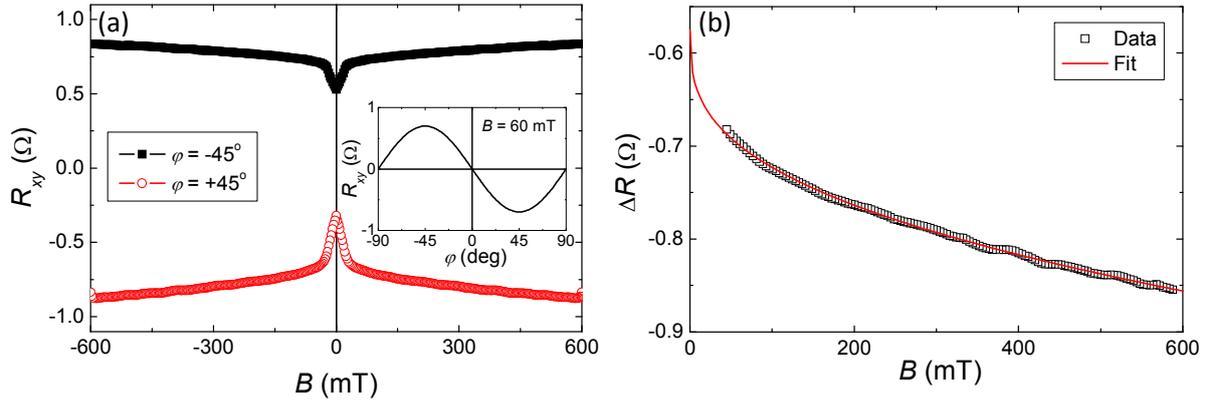

Fig. S1 (a) Transverse resistance $R_{xy}$ measured at $T$ = 150 K as a function of the external magnetic field $B$ applied at $\varphi$ = -45° (closed squares) and $\varphi$ = +45° (open circles). The inset shows $R_{xy}$ measured at $T$ = 150 K as a function of $\varphi$, the angle of the applied $B$, for $B$ = 60 mT. (b) Amplitude $\Delta R$ of the AMR at $T$ = 150 K as a function of $B$ obtained from [$R_{xy}(\varphi$ = -45°) - $R_{xy}(\varphi$ = +45°)]/2 (open squares) and corresponding fitting (solid line) using the expression $\Delta R(B) = \Delta R_0 + a'B + b'B^{1/3}$.